\documentclass[twocolumn,twoside,10pt]{asme2e}
\usepackage{graphicx}
\usepackage{cite}            
\usepackage{amssymb}
\usepackage{ifthen}
\usepackage[noprefix]{nomencl}
\usepackage{ifpdf}\ifpdf 
\usepackage[colorlinks,linktocpage,pdftex,hyperfigures]{hyperref} \fi
\usepackage[switch]{lineno}

\renewcommand{\nomgroup}[1]{\medskip
\ifthenelse{\equal{#1}{B}}{\item[\textbf{Abbreviations}]}{ \ifthenelse{\equal{#1}{G}}{\item[\textbf{Greek
symbols}]}{ \ifthenelse{\equal{#1}{S}}{\item[\textbf{Subscripts}]}{
\ifthenelse{\equal{#1}{P}}{\item[\textbf{Superscripts}]}{}}}}}
 \makenomenclature

\title{A dynamic film  model of the Pulsating Heat Pipe}

\author{Vadim S. Nikolayev
    \affiliation{ESEME, Service des Basses Temp\'eratures, INAC/CEA-Grenoble, France\\and\\
    ESEME, PMMH-ESPCI, 10, rue Vauquelin, 75231 Paris Cedex 5, France\\
    Email: vadim.nikolayev@espci.fr}}

\begin{document}
\maketitle


\section*{Abstract}\begin{abstract}
This paper deals with the numerical modelling of  the pulsating heat pipe (PHP) and is based on the film
evaporation/condensation model recently applied to the single bubble PHP (Das et al., 2010). The described
numerical code can treat the PHP of arbitrary number of bubbles and branches. Several phenomena that occur
inside the PHP are taken into account: coalescence of liquid plugs, film junction or rupture, etc. The model
reproduces some of the experimentally observed regimes of functioning of the PHP like chaotic or intermittent
oscillations of large amplitude. Some results on the PHP heat transfer are discussed.
\end{abstract}

\noindent\textit{Keywords: oscillating, pulsating, heat pipe, simulation}

\section{Introduction}\label{sec:intro}

The pulsating (or oscillating) heat pipe (PHP) is a long capillary tube bent into many branches and partially
filled with a two-phase, usually single component, working fluid \cite{akachi}. The tube is simple, a wick
structure is not required. The fluid spontaneously forms multiple vapor bubbles separated by liquid plugs inside
the tube. Evaporation of liquid in the hot (evaporator) sections and subsequent condensation in the cold
(condenser) sections creates oscillations of the bubble-plug structure. These oscillations are very important
because they lead to a substantial increase of the heat transfer rate in comparison with other types of heat
pipes \cite{Vasiliev}. In addition to the latent heat transfer characteristic for them, the sensible heat
transfer is important in PHP. While sweeping a section belonging to evaporator, a liquid plug accumulates the
heat, which is then transferred to the condenser section when the plug penetrates there.

Because of their simplicity and high performance, PHPs are often considered as highly promising. Their
industrial application is however limited because their functioning is non-stationary and thus difficult to be
controlled. During the last decade, researchers have extensively studied PHPs \cite{zh_faghri_rev}. Tong et al.
\cite{tong}, Miyazki and Arikawa \cite{miyazaki}, Khandekar et al. \cite{Khan03}, Xu et al. \cite{xu}, Gi et al.
\cite{gi} and Inoue et al. \cite{Inoue10} have carried out flow visualization studies with several working
fluids. These experiments confirmed the existence of self-sustained thermally driven oscillations in PHPs.
Several experimental groups \cite{khan1,Li08,Ma08,Maydanik09,Lips10} performed experiments with different tube
diameters, configurations, orientations and filling ratios and studied the thermal performance of PHPs in
different conditions. However, the functioning of PHPs is not completely understood. A complicated interplay of
different hydrodynamic and phase-exchange phenomena needs to be accounted for. The experimental studies of the
background physical phenomena that cause the instability or are responsible for the PHP behavior are only a few
\cite{Lips10,IJHMT10}.

There are several modeling approaches available in the literature. Shafii et al. \cite{shafii1} initiated the
modeling approaches for multi-branch PHPs. The evaporation-condensation mass exchange was accounted for with the
temperature difference terms $\propto (T_{wall}-T_i)$ where $T_{wall}$ was either $T_e$ or $T_c$ depending on
the bubble location. The problem was solved numerically with the explicit Euler scheme. Periodical (nearly
sinusoidal) oscillations appeared after a transient. Their amplitude was small: the displacement amplitude did
not exceed the evaporator size. It was concluded that the heat is transferred mainly via the sensible heat
transfer; the latent heat transfer was an order of value smaller. The same model has been used later by another
team \cite{Sakul}.

It is well known from the analysis of the conventional heat pipes that in reality, most of evaporation in the
evaporator occurs through the liquid films that might cover only a part of the heated surface. Dobson
\cite{dobson1,dobson2} introduced a lumped meniscus model where the films are considered to be of constant
thickness $\delta_f$ but of varying length. Apart from the film introduction, the model was similar to
\cite{shafii1}. Single bubble PHP with an open end was considered. The oscillations were unstable and
consisted of a nearly periodical pattern which began with a strong displacement during which the meniscus
penetrated into the evaporator. This initiated high frequency declining to zero oscillations around an
average position situated in the condenser. Das et al. \cite{IJHMT10} attempted to reproduce the results of
Dobson with his model for the same parameters. They obtained only small amplitude periodical oscillations
during which the meniscus never penetrated into the evaporator. They attributed the disagreement to the poor
stability of the numerical algorithm (explicit Euler) used by Dobson. The 4th order Runge-Kutta method, well
known to be stable, was used in \cite{IJHMT10}. A rigorous analytic analysis of a simplified version of the
model, where the evaporation-condensation dynamics is modeled with the $(T_{wall}-T_i)$ term, has been also
carried out by Das et al. An analytic expression giving the condition, under which the self-sustained
oscillations appear, was obtained. It was shown that such a model leads necessarily to small amplitude
oscillations.

A 2D model of the single-bubble PHP was considered by Zhang and Faghri \cite{zhang-faghri}. A conceptual
difference with the previous approaches concerned the vapor equation of state. Instead of the ideal gas
model, the vapor was considered to be at saturation temperature $T_{sat}$ corresponding to its pressure $P$.
Small amplitude periodic nearly harmonic oscillations were obtained. Holley and Faghri \cite{Holley05}
applied the same assumption to the PHP with spatially varying diameter.

Das et al.  introduced the film evaporation-condensation model. The film is introduced similarly to
\cite{dobson1,dobson2}. The vapor mass exchange is assumed to be limited by the heat conduction in the film like
in the work \cite{zhang-faghri}. This leads to the mass exchange rate
$\lambda_l[T_{wall}-T_{sat}(P)]/(\delta_fh_{lv})$ where $T_{sat}(P)$ is the gas-liquid interface temperature.
This approach is different from all previous approaches because a strong temperature gradient is assumed to
exist in the vapor so that the temperature $T$ of its bulk is allowed to be different from $T_{sat}(P)$. The
validity of this assumption is checked in \cite{IJHMT10} \emph{a posteriori}. The simulations have shown that
most of the time $T>T_{sat}(P)$. The same form of the mass exchange term has been used recently by Senjaya et
al. \cite{Senjaya10}. They however did not use the variable films. The film evaporation-condensation model was
validated against the single branch experiment \cite{IJHMT10}. It reproduces oscillations the amplitude of which
might be larger than the size of evaporator. The purpose of the present article is to apply it to the
multi-bubble PHP. The closed loop PHP will be considered. The model is however can be applied also to the
unlooped PHP.

 \nomenclature[a]{$Nu$}{liquid Nusselt number}
 \nomenclature[a]{$Q$}{heat exchange rate [W]}
 \nomenclature[s]{$e$}{evaporator}
 \nomenclature[s]{$f$}{liquid film}
\nomenclature[s]{$p$}{PHP spatial period} \nomenclature[p]{$sens$}{sensible} \nomenclature[p]{$lat$}{latent}
\nomenclature[g]{$\Delta$}{difference}
 \nomenclature[p]{$l$}{left}
 \nomenclature[p]{$r$}{right}
 \nomenclature[p]{$s$}{$r$ or $l$}
 \nomenclature[s]{$s$}{$e$ or $c$}
 \nomenclature[p]{$r$}{right}
  \nomenclature[s]{$m$}{meniscus}
  \nomenclature[p]{$k$}{index}
  \nomenclature[s]{$j$}{film junction}
  \nomenclature[s]{$l$}{liquid}
  \nomenclature[s]{$b$}{branch}
  \nomenclature[s]{$a$}{adiabatic}
 \nomenclature[s]{$t$}{total}
 \nomenclature[s]{$c$}{condenser}
 \nomenclature[a]{$h_{lv}$}{latent heat [J/kg]}
 \nomenclature[s]{$v$}{vapor}
 \nomenclature[s]{$wall$}{internal tube wall}
 \nomenclature[s]{$i$}{bubble or plug identifier}
 \nomenclature[a]{$n$}{sequential branch number}
\nomenclature[a]{$M$}{total number of bubbles or plugs} \nomenclature[a]{$N$}{total number}
\nomenclature[a]{$T$}{temperature ($T_i$: of vapor) [K]}
 \nomenclature[a]{$X$}{absolute position at the $x$ axis [m]}
 \nomenclature[a]{$g$}{gravity acceleration [m$^2$/s]}
\nomenclature[a]{$L$}{length [m]}
 \nomenclature[a]{$m$}{mass (of vapor by default) [kg]}
 \nomenclature[a]{$V$}{velocity ($V_i$: of liquid plug) [m/s]}
 \nomenclature[a]{$K$}{friction coefficient}
 \nomenclature[a]{$c_{vv}$}{vapor specific heat at constant volume [J/(kg$\cdot$K)]}
 \nomenclature[a]{$U$}{heat transfer coefficient (of film transfer if no indices) [W/(K$\cdot$m$^2$)]}
 \nomenclature[g]{$\lambda$}{heat conductivity [W/(m$\cdot$K)]}
 \nomenclature[a]{$F$}{viscous friction force [N]}
 \nomenclature[a]{$G$}{gravity force [N]}
 \nomenclature[a]{$P$}{vapor pressure [Pa]}
 \nomenclature[a]{$t$}{time [s]}
\nomenclature[a]{$Re$}{liquid Reynolds number}
 \nomenclature[g]{$\delta_f$}{liquid film thickness [m]}
 \nomenclature[a]{$d$}{tube diameter [m]}
 \nomenclature[a]{$S$}{tube section area [m$^2$]}
\nomenclature[a]{$R_v$}{vapor gas constant [J/(kg$\cdot$K)]}
 \nomenclature[s]{$sat$}{at saturation}
 \nomenclature[g]{$\rho$}{density [kg/m$^3$]}
 \nomenclature[g]{$\nu$}{liquid kinematic viscosity [m$^2$/s]}
 \nomenclature[a]{$D$}{liquid heat diffusivity [m$^2$/s]}
\nomenclature[g]{$\gamma$}{coefficient in Eq. \ref{U}} \nomenclature[g]{$\phi$}{volume fraction of liquid in
PHP}

\section{Problem statement}
\begin{figure}[htb]
\centering
\includegraphics[width=\columnwidth]{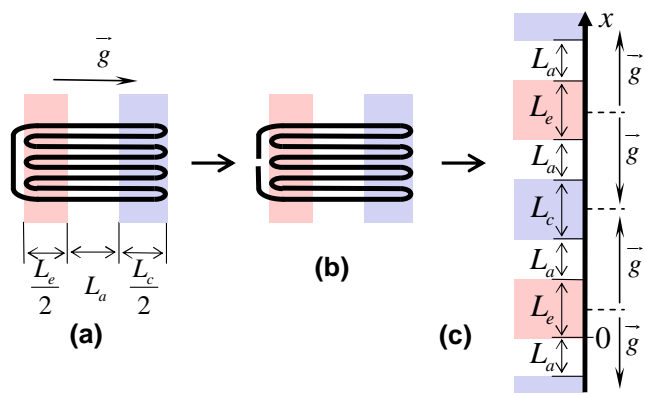}
\caption{(Color online) (a) Sketch of the closed loop PHP with the default gravity direction. Evaporator is
to the left. (b,c) Topological transformations of the tube. (b) Unlooping. (c) Unbending and projection to
the $x$ axis. The splitting of the $x$ axis to branches is shown.} \label{Simu}\end{figure} Like in
\cite{shafii1,Sakul}, the PHP meandering tube is projected to a straight axis $x$ so that it consists of
periodic sequence of different domains corresponding to the PHP sections (Fig. \ref{Simu}). One PHP spatial
``period'' of the length $L_p=2L_a+L_c+L_e$ is assumed to contain the sections in the following order:
evaporator, adiabatic, condenser, adiabatic. The point $x=0$ is assumed to coincide with the beginning of an
evaporator. The PHP branch is a half of a period ($L_b=0.5L_p$) beginning in the middle of a condenser or an
evaporator. $L_t=N_pL_p$ is the total PHP length. Each bubble is identified by the index $i$. The neighboring
from the right side liquid plug is denoted by the same index. The total number $M$ of bubbles may change in
time.

Unlike \cite{shafii1}, the axis and the periodical pattern of sections on it are continued to infinity in
both directions. At $t=0$, the bubbles are positioned at the axis and may move at $t>0$ along the infinite
axis $x$ as far as needed. This means that the $x$ value itself does not have any significance; only the
relative positions are meaningful; they are determined with the remainder operator defined as $y\bmod
z=y-z\,\mbox{int}(y/z)\geq 0$ where $\mbox{int}(z)$ means the integer part of $z$, i.e. the largest integer
smaller than $z$. E. g. $x$ belongs to evaporator if $x\bmod L_p\le L_e$. Since the PHP loop is closed, each
point $x$ is equivalent to the point $L_t+x$. Such an approach is convenient because it simplifies the
management of any kind of bubble motion, in particular their unidirectional circulation. Within such a
description, the coordinate $X_i^l$ of the left meniscus of the bubble $i$ is always smaller than that of its
right meniscus $X_i^r$ and the bubble order does not change during their motion. Note that the coordinate
$X_M^r$ of the left end of the last liquid plug is larger than that of its right end $X_1^l$.

The constant temperatures $T_e$ and $T_c$ are imposed at the inner walls of the evaporator and condenser.

\subsection{Film dynamics in evaporators}\label{film_d}
\begin{figure}[htb]
\centering
\includegraphics[width=\columnwidth]{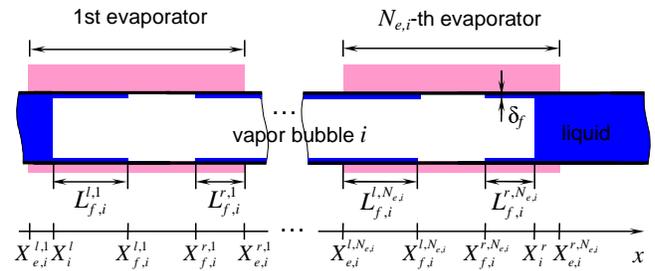}
\caption{(Color online) Vapor bubble $i$ that extends over $N_{e,i}$ evaporators and the liquid films inside
it.} \label{bubble}\end{figure} According to the film evaporation/condensation model \cite{IJHMT10}, the film
length may vary because of two reasons: (i) evaporation or condensation and (ii) film deposition
(Landau-Levich phenomenon) or film ``eating up'' during receding or advancing of the liquid meniscus,
respectively. We generalize this model here to the case of multiple branches. The liquid film is assumed to
always cover the inner walls of the tube (i.e. to be continuous) in the condenser and adiabatic sections when
the whole bubbles or their parts are located inside them. The film may be partially or completely evaporated
in the evaporator. The films are assumed to have a homogeneous thickness $\delta_f$ in all the sections.

Let us consider the $i$-th bubble that may extend over several PHP bends or, equivalently, evaporators (Fig.
\ref{bubble}). A number $N_{e,i}$ of pairs of films per bubble (left+right each) coincides with the number
$\textrm{int}({X_i^r}/{L_p})-\textrm{int}[(X_i^l-L_e)/{L_p}]$ of evaporators to which at least a part of the
bubble belongs if the latter quantity is nonzero. In the opposite case, $N_{e,i}=1$ to describe the continuous
film in the adiabatic and condenser sections. The evaporators are counted left-to-right by the index $k=1\dots
N_{e,i}$. The (non-negative) lengths of the left and right films of the bubble in the $k$-th evaporator are
denoted $L_{f,i}^{l,k}$ and $L_{f,i}^{r,k}$, respectively. When the film in the evaporator is not continuous,
their dynamics is described by the equations (cf. Eq. (26) of \cite{IJHMT10})
\begin{eqnarray}
\dot{L}_{f,i}^{l,k}&=&\left\{ \begin{array}{ll}
0 & \textrm{if $L_{f,i}^{l,k}= 0$, $\dot{X}_i^l>0$ and} \\&X_i^l\in k\mbox{-th evaporator},\\
-\dot m_{fe,i}^{l,k}/(\rho_l\pi d\delta_f)  & \textrm{if $X_i^l\notin k$-th evaporator},\\
-\dot m_{fe,i}^{l,k}/(\rho_l\pi d\delta_f)-\dot{X}_i^l  & \mbox{otherwise,}
\end{array} \right.\label{Lfl}\\
\dot{L}_{f,i}^{r,k}&=&\left\{ \begin{array}{ll}
0 & \textrm{if $L_{f,i}^{r,k}= 0$, $\dot{X}_i^r<0$ and}\\ &X_i^r\in k\mbox{-th evaporator},\\
-\dot m_{fe,i}^{r,k}/(\rho_l\pi d\delta_f)  & \textrm{if $X_i^r\notin k$-th evaporator},\\
-\dot m_{fe,i}^{r,k}/(\rho_l\pi d\delta_f)+\dot{X}_i^r  & \mbox{otherwise},
\end{array} \right.\label{Lfr}
\end{eqnarray}
where dot means the time derivative. Evidently, the condition ``$X_i^s\in k$-th evaporator'' may be satisfied
only for $k=1$ if $s=l$ and for $k=N_{e,i}$ if $s=r$. The mass evaporation rate at $k$-th evaporator $\dot
m_{fe,i}^{s,k}$ from the film $s\in\{r,l\}$ is defined by the interfacial heat balance equation
\begin{equation}\label{mfei}
    h_{lv}\dot m_{fe,i}^{s,k}=U\pi dL_{f,i}^{s,k}[T_e-T_{sat}(P_i)].
\end{equation}
As discussed in the Introduction, the heat transfer coefficient is defined by the expression
\begin{equation}\label{U}
    U=\gamma \lambda_l/\delta_f,
\end{equation}
where $\gamma\lesssim 1$ is a coefficient accounting for the spatial variation of the film thickness that exists
in reality. The meaning of Eqs. (\ref{Lfl}, \ref{Lfr}) is simple:  the film length may decrease due to
evaporation or due to the meniscus advancing in the direction of the film edge (``eating up" of the liquid
film). The film length increases when the film is left during the receding motion of the meniscus (Landau-Levich
film deposition). It is assumed that the triple contact line (i.e. film edge) is pinned and does not recede in
the absence of evaporation.

The coordinates of the left and right film edges in each of $N_{e,i}$ evaporators ($X_{f,i}^{l,k}$ and
$X_{f,i}^{r,k}$, respectively) may coincide with one of the ends of the evaporator or with a position of a
meniscus if the corresponding film length is zero. Provided $X_i^l,X_i^r,L_{f,i}^{l,k},L_{f,i}^{r,k}$ are known,
$X_{f,i}^{l,k}$ and $X_{f,i}^{r,k}$ can be determined from the following reasoning.

When no part of the bubble situates in evaporators, $N_{e,i}=1$ and the film is continuous as mentioned
above. This is equivalent to a junction of the left and right films at a point (denoted $X_j^1$) that needs
to be chosen. This choice is of importance because in the simulation it defines the point of the film rupture
that may occur. Therefore $X_j^1=X_{f,i}^{l,1}=X_{f,i}^{r,1}$ is assumed to coincide with the left or right
meniscus (i.e with $X_i^l$ or $X_i^r$), correspondingly to the direction of the bubble motion so that the
evolution of the film edges exhibits no discontinuity when the Landau-Levich film begins to be deposed
according to (\ref{Lfl}, \ref{Lfr}).

When at least a part of the bubble situates in evaporators, one applies the following expressions to each of
them ($k=1\dots N_{e,i}$):
\begin{eqnarray}
X_{f,i}^{l,k} &=& \left\{ \begin{array}{ll}
X_i^l  & \textrm{if }X_i^l\in k\textrm{-th evaporator} \\
X_{e,i}^{l,k}& \mbox{otherwise}
\end{array} \right\}+L_{f,i}^{l,k},\label{Xfl}\\
X_{f,i}^{r,k} &=& \left\{ \begin{array}{ll}
X_i^r  &\textrm{if } X_i^r\in k\textrm{-th evaporator} \\
X_{e,i}^{r,k} & \mbox{otherwise}
\end{array} \right\}-L_{f,i}^{r,k}.\label{Xfr}
\end{eqnarray}
Here
\begin{equation}\label{Xeb}
X_{e,i}^{l,k}=L_p\left[\textrm{int}\left(\frac{X_i^l-L_e}{L_p}\right)+k\right]
\end{equation}
is the left edge of the $k$-th evaporator and $X_{e,i}^{r,k}=X_{e,i}^{l,k}+L_e$ is its right edge.

At some occasion, Eqs. (\ref{Xfl}, \ref{Xfr}) may result in $X_{f,i}^{l,k}\geq X_{f,i}^{r,k}$ (i.e in a small
film overlap allowed in the numerical calculation where the time steps are discrete). This signifies an
appearance of a continuous liquid film. The point $X_j^k$ (where the film will disrupt should the film rupture
occur) needs to be defined. It is assumed to coincide with the point of film junction where the film is thinner
there and likely to be evaporated quicker than elsewhere. Once $X_j^k$ is defined, $X_{f,i}^{r,k}$ and
$X_{f,i}^{l,k}$ are reassigned to $X_j^k$. Next, $L_{f,i}^{r,k}$ and $L_{f,i}^{l,k}$ are recalculated with Eqs.
(\ref{Xfl}, \ref{Xfr}).

\subsection{Remaining vapor bubble governing equations}
Condensation occurs to the film that surrounds a bubble portion $L_{c,i}$ located in the condenser. Similarly to
the evaporation description, the film condensation rate $m_{fc,i}$ is defined by the expression
\begin{equation}\label{mfci}
h_{lv}\dot m_{fc,i}=U\pi dL_{c,i}[T_{sat}(P_i)-T_c].
\end{equation}

Although much weaker than at the film interface, the mass exchange occurs at the remaining meniscus part (its
portion of the size $L_m\ll d$ adjacent to the tube wall) and exists even if the film is evaporated
completely. The heat balance on the meniscus depends on whether the meniscus $s\in\{r,l\}$ situates inside
the evaporator or the condenser,
\begin{eqnarray}
h_{lv}\dot m_{me,i}^s&=&U_m\pi dL^s_{me,i}[T_e-T_{sat}(P_i)],\\
h_{lv}\dot m_{mc,i}^s&=&U_m\pi dL^s_{mc,i}[T_{sat}(P_i)-T_c],
\end{eqnarray}
where $U_m<U$ and
\begin{eqnarray}
L^s_{me,i} = \left\{ \begin{array}{ll}
L_m  & \mbox{if}\; X_i^s\in\textrm{ evaporator},  \\
0 & \mbox{otherwise},
\end{array} \right.\\
L^s_{mc,i} = \left\{ \begin{array}{ll}
L_m  & \mbox{if}\; X_i^s\in\textrm{ condenser},  \\
0 & \mbox{otherwise},
\end{array} \right..\\
\end{eqnarray}
The total bubble mass change rate $\dot m_i$ can be expressed as
\begin{equation}
\dot{m}_i=\sum_{s=r,l}\left[\sum_{k=1}^{N_{e,i}}\dot m_{fe,i}^{s,k}+\dot{m}_{me,i}^s-\dot m_{mc,i}^s\right]-\dot
m_{fc,i}.\label{mv}
\end{equation}

The energy equation of the $i$-th bubble is \cite{shafii1,Comment11}
\begin{equation}\label{en}
m_ic_{vv}\dot{T}_i=\dot m_iR_vT_i+Q^{sens}_i-P_iS(\dot{X}_i^r-\dot{X}_i^l),
\end{equation}
where the sensible heat exchange with the dry evaporator walls is described by the expressions
\begin{equation}
    Q^{sens}_i=U_v\pi d L^{sens}_i(T_e-T_i),\label{qsens}
\end{equation}
where $U_v=2\lambda_v/d$ (it is assumed that the boundary layer in the vapor is the tube radius) and
\begin{equation}
    L^{sens}_i=\sum_{k=1}^{N_{e,i}}(X_{f,i}^{r,k}-X_{f,i}^{l,k})\label{Lsens}.
\end{equation}
The vapor equation of state is approximated with the ideal gas equation
\begin{equation}\label{EOS}
    P_i=\frac{m_iR_vT_i}{S(X_i^r-X_i^l)}.
\end{equation}
Note that the ideal gas assumption is used \cite{shafii1,Comment11} while deriving (\ref{en}).

\subsection{Liquid plug governing equations}
The $i$-th liquid plug, i.e. that to the right of the $i$-th bubble, has the mass
\begin{equation}\label{mldef}
m_{l,i}=\rho_lS(X_{i+1}^l-X_i^r),
\end{equation}
where the index $i+1$ denotes the bubble to the right of the $i$-th bubble. The velocity $V_i$ of the plug's
center of mass is
\begin{equation}\label{Vi}
    V_i=0.5(\dot{X}_{i+1}^l+\dot{X}_i^r).
\end{equation}
It obeys the momentum equation
\begin{equation}
\frac{d(V_{i}m_{l,i})}{dt}=(P_{i}-P_{i+1})S-F_i\mbox{ sign}(V_i)+G_i, \label{Vl}
\end{equation}
where $G_i$ is the gravity term discussed below. The viscous friction force $F_i$ is defined by the
expression corresponding to the single phase friction \cite{dobson1}
\begin{eqnarray}
F_i&=&\frac{1}{2} K d\rho_l \pi (X_{i+1}^l-X_i^r) V_i^2\label{Ft}\\
K &=& \left\{ \begin{array}{ll}
16 & Re<1 \\
16/Re & 1\leq Re<1180\\
0.078Re^{-0.25} & Re \geq 1180
\end{array} \right.\nonumber,\\Re&=&|V_i| d/\nu.
\end{eqnarray}
As in the previous modelling approaches, an additional contribution of the bends to $F_i$ is neglected.

The liquid is assumed to be incompressible. This implies that the liquid plug volume may vary because of only
two reasons: (i) phase change at its menisci and (ii) liquid film deposition or, on the contrary, ``eating up".
The velocity of the plug ends relative to its center of mass is thus non-zero and is defined by the change in
the plug volume. This condition leads to the following condition of liquid mass balance in the plug:
\begin{eqnarray}
&&\dot m_{l,i}=\dot m_{me,i+1}^l-\dot m_{mc,i+1}^l+\dot m_{me,i}^r-\dot m_{mc,i}^r+\pi
d\delta_f\rho_l\nonumber\\&&\left[\left\{
\begin{array}{ll}
0 & \mbox{if}\; X_{i+1}^l\in\textrm{ evaporator, }\dot{X}_{i+1}^l>0,\; \mbox{and}\; L_{f,i+1}^{l,1}=0,\\
\dot{X}_{i+1}^l & \mbox{otherwise}.
\end{array} \right\}\right.\nonumber\\&&-\left.\left\{ \begin{array}{ll}
0 & \mbox{if}\; X_i^r\in\textrm{ evaporator, }\dot{X}_i^r<0\; \mbox{and}\; L_{f,i}^{r,N_{e,i}}=0,\\
\dot{X}_i^r  & \mbox{otherwise},\end{array}\right\} \right]\label{dml}
\end{eqnarray}
The upper options in both braces correspond to the meniscus advancement over the dry evaporator, and the lower options, to the film deposition or
``eating up". The equations for $\dot{X}_i^l,\dot{X}_i^r$ need to be obtained from (\ref{mldef}, \ref{Vi},
\ref{dml}). It is evident that $\dot{X}_{i+1}^l\approx\dot{X}_i^r\approx V_i$ within quite small terms of the
order $\delta_f/d$ and $\rho_v/\rho_l$ that describe the liquid volume variation. For this reason, $V_i$ can
be used in the conditional clauses of (\ref{dml}) instead of $\dot{X}_{i+1}^l,\dot{X}_i^r$. The set of
equations (\ref{mldef}, \ref{Vi}, \ref{dml}) becomes linear and can be solved straightforwardly for
$\dot{X}_i^l,\dot{X}_i^r$. We do not however write their explicit expressions here because they are
cumbersome.

The liquid volume variation was neglected in previous works. It is introduced here to provide the
conservation of the total fluid mass in the PHP. A small error that arises when the conservation is violated
accumulates and may become important at large simulation times.

\subsubsection{Gravity term}

The gravity sign is constant along each PHP branch but alters between branches, see Fig. \ref{Simu}c. The
default gravity direction coincides with the $x$ axis direction within the branch number $n=0$ that starts at
$x=0.5L_e$. The gravity sign corresponding to the branch number $n$ is thus $(-1)^n$. The branch number at a
given $x$ is
\begin{equation}\label{brn}
    n_b(x)=\mbox{int}\left(\frac{x-0.5L_e}{L_b}\right).
\end{equation}
The branch numbers of the liquid plug left and right ends are therefore $n^l_i=n_b(X_i^r)$ and
$n^r_i=n_b(X_{i+1}^l)$, respectively. The gravity force reads
\begin{eqnarray}
&&G_i=\rho_lSg\Bigg\{(-1)^{n^r_i}\left[\left(X_{i+1}^l-\frac{1}{2}L_e\right)\bmod
L_b-\frac{1}{2}L_b\right]-\nonumber\\&&(-1)^{n^l_i}\left[\left(X_i^r-\frac{1}{2}L_e\right)\bmod
L_b-\frac{1}{2}L_b\right]\Bigg\}.\label{G1}
\end{eqnarray}
It is evident that for $n^l_i=n^r_i$ (i.e. when the entire plug belongs to one branch),
\begin{equation}\label{G0}
    G_i=m_{l,i}g(-1)^{n^l_i}.
\end{equation}
The inclination angle $\theta$ of the PHP with respect to the vertical direction can be simulated by
replacing $g$ by $g\sin\theta$. $\theta=\pi$ can be used to describe the opposite (condenser on the top) PHP
position.

\subsubsection{Heat diffusion in liquid}

The temperature distribution in the liquid plug $T_{l,i}=T_{l,i}(x,t)$ where $x\in(X_i^r,X_{i+1}^l)$ is governed
by the heat diffusion equation \cite{shafii1},
\begin{equation}\label{hd}
\frac{\partial T_{l,i}}{\partial t}=D \frac{\partial^2 T_{l,i}}{\partial x^2}+D\frac{4
Nu}{d^2}(T_{wall}-T_{l,i}).
\end{equation}
The last term accounts for the heat transfer with the tube wall \cite{shafii1}. The Nusselt number depends on
the plug velocity and length. The expressions for $Nu$ for different ranges of $Re$ are taken from \cite{Bejan}.
$T_{wall}$ is $T_e$ or $T_c$ depending on where $x$ situates. The last term is absent at all if $x$ belongs to
the adiabatic section.

The boundary conditions for eq. (\ref{hd}) are given at the menisci:
\begin{eqnarray}\label{bc} T_{l,i}(X_i^r)&=&T_{sat}(P_i),\\
T_{l,i}(X_{i+1}^l)&=&T_{sat}(P_{i+1}).
\end{eqnarray}
Note that all previous equations of the model are independent of $T_l$ because $T_e$ and $T_c$ are imposed
and independent of the heat load. Eq. (\ref{hd}) can thus be solved \emph{after} the calculation of the PHP
dynamics. In our numerical code it is however solved together with all other equations so that the boundary
conditions where $T_e$ and $T_c$ depend on the heat load could be easily implemented in the future.

\subsection{Heat exchange rates}

The instantaneous sensible heat power taken by the $i$-th bubble-plug pair from the evaporator is calculated
with the following expression:
\begin{equation}\label{Qsensei}
    Q^{sens}_{e,i}=2\pi\lambda_l\int Nu (T_e-T_{l,i})\textrm{d}x+Q^{sens}_i,
\end{equation}
where the integration is performed over the portion of the liquid plug located in the evaporator. The heat
power given to the condenser is calculated accordingly,
\begin{equation}\label{Qsensci}
    Q^{sens}_{c,i}=2\pi\lambda_l\int Nu (T_{l,i}-T_c)\textrm{d}x,
\end{equation}
where the integration is performed over the portion of the liquid plug located in the condenser. The
instantaneous latent heat power taken by the $i$-th bubble-plug pair from the evaporator is
\begin{equation}\label{Qlatei}
    Q^{lat}_{e,i}=h_{lv}\sum_{s=r,l}\sum_{k=1}^{N_{e,i}}\dot m_{fe,i}^{s,k}.
\end{equation}
That given to condenser is
\begin{equation}\label{Qlatci}
    Q^{lat}_{c,i}=h_{lv}\dot m_{fc,i}.
\end{equation}

The instantaneous heat power is a sum of the corresponding terms over all bubbles of the PHP,
\begin{equation}\label{Q}
   Q^k_s=\sum_i^M Q^k_{s,i},
\end{equation}
where $k\in\{sens,lat\}$ and $s\in\{e,c\}$. The total heat power reads
\begin{equation}\label{Qt}
    Q_s=Q^{lat}_s+Q^{sens}_s.
\end{equation}
In the stationary regime, the condition
\begin{equation}\label{Qcond}
    \langle Q_c\rangle=\langle Q_e\rangle,
\end{equation}
where the angle brackets mean time average should be valid for long enough averaging times: the amount of heat
taken from the evaporator should be equal to that given to the condenser.

\subsection{Bubble-plug events}\label{events}

There are several kinds of events that can change the bubble-plug morphology. They cause a change of the
equations to be solved on the next time step. One of such events, the film junction, was described in sec.
\ref{film_d}. It changed equations but conserved their number. In this section, we rather discuss the events
that change the number of both differential equations and unknowns.

First, it is the vapor bubble recondensation. It occurs when a moving liquid plug overtakes another plug. The
vapor pressure grows and fast condensation occurs. A bubble located between two plugs completely disappears
and a new long plug forms. Its mass is a sum of the masses of the parent plugs and its velocity is determined
from the momentum conservation. On the next time step, the number of bubble-plug pairs drops by one and the
number of equations reduces accordingly.

Another event met very often is a change in the number of liquid films. Such an event occurs when a bubble
penetrates into extra evaporator or, on the contrary, withdraws from it. $N_{e,i}$ number then changes which
means that the number of differential equations (\ref{Lfl},\ref{Lfr}) changes too.

Other yet non implemented events include the vapor bubble creation by boiling or the complete liquid plug
evaporation.

\section{Numerical implementation}

The spatial integration of eq. (\ref{hd}) is performed at each time step and needs to be discussed first.

\subsection{Spatial integration of the heat diffusion equation}

The $i$-th plug is divided to $N_{l,i}+2$ finite elements $\Delta x^k_i$. All (except of two ending) elements
are of the same length; two ending elements are half-length. The element length varies slightly from plug to
plug to keep the total number of elements integer. The node points $X^k_i$ are in the centers of the internal
$N_{l,i}$ elements. The node temperatures are denoted $T_{l,i}^k$. The temperature values at $X^0_i=X^r_i$ and
$X^{N_{l,i}+1}_i=X^l_{i+1}$ are given by the boundary conditions (\ref{bc}), so that there are $N_{l,i}$ unknown
temperatures per plug. The finite volume method \cite{Patankar} is used and eq. (\ref{hd}) is integrated over
each element. This results in the following discrete analog of equation (\ref{hd}) written for $k=1\dots
N_{l,i}$,
\begin{eqnarray}\label{hdd}
\left.\frac{\partial T_{l,i}}{\partial t}\right|^k = \frac{2D}{\Delta
x^k_i}\left(\frac{T_{l,i}^{k+1}-T_{l,i}^k}{\Delta x^{k+1}_i+\Delta x^k_i}+\frac{T_{l,i}^{k}-T_{l,i}^{k-1}}{\Delta
x^k_i+\Delta x^{k-1}_i}\right)\nonumber\\ + D\frac{4 Nu}{d^2}\left(T_{wall}-T_{l,i}^k\right).
\end{eqnarray}

\subsection{Data structure}\label{Data}

The code is object oriented and is written with C++. All variables are dynamically allocated. The code can thus
deal with a PHP with arbitrary PHP geometry and time varying number of plugs, bubbles, and liquid films. The
code makes use of the Microsoft Foundation Class (MFC) library to take advantage of the serialization (saving
and restoring to/from data files) of the objects of the unknown in advance size. Each vapor bubble or liquid
plug is implemented as a C++ object and encapsulates a number of scalar and vector ``member'' variables proper
to each of them. The $i$-th bubble scalar members include $T_i, m_i, P_i,
X_i^l,X_i^r,N_{e,i},Q^{lat}_{c,i},Q^{lat}_{e,i}$, a unique bubble identification number, and a pointer to the
neighboring (from the right) liquid plug. The pointer is simply a computer memory address where the target
object is located. The pointer may have null value to indicate the plug absence, which is useful to simulate the
last bubble in the unlooped PHP. In the present article, only the closed loop PHP is considered so that each
bubble has a plug. The vector member variables of the bubble include the liquid film lengths and edge
coordinates. They are allocated dynamically and so that their length $N_{e,i}$ may vary. The scalar plug
variables include $V_i,m_{l,i},N_{l,i},Q^{sens}_{c,i},Q^{sens}_{e,i}$, and a pointer to its (left neighbor)
bubble. The vector members of a plug are $\vec{\Delta x}_i$, $\vec{X}_i$, and $\vec{T}_{l,i}$.

\begin{figure}[htb]
\centering
\includegraphics[width=0.9\columnwidth,clip]{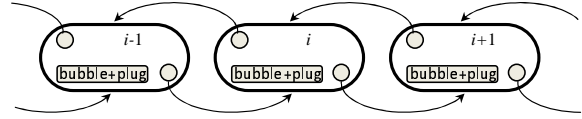}
\caption{Computer representation of the instantaneous state of the PHP as a doubly connected list. Each its node
(an oval) contains the state of a bubble-plug pair and two pointers (circles) to the previous and to the next
nodes.} \label{DList}
\end{figure}
In the remaining part of this section, the term PHP is used to denote the instantaneous state of all its
bubbles and plugs. PHP is implemented as a doubly connected list (Fig. \ref{DList}). The doubly connected
list is an array, the $i$-th node of which contains the data (of a bubble and its plug) and the pointers to
both previous and next nodes. Note that in this case $i$ is an identifier of the node rather than its
sequential number. The previous and the next nodes are denoted as $i-1$ and $i+1$ respectively just for the
sake of illustration. The double connectivity allows fast access to these variables. Unlike the conventional
(e.g. FORTRAN) arrays, such data structures can contain objects of different and variable length. This is
convenient for several reasons. One of them is that the objects corresponding to plugs may contain the
vectors of different and time-variable sizes $N_{l,i}$. The lists are convenient for another reason. Unlike
the standard arrays, the list nodes are not necessarily written continuously into the computer memory.
Therefore, it is very easy to suppress or, on the contrary, add a node somewhere in the middle of the PHP
(which corresponds to the bubble recondensation or nucleation, respectively) without modifying the whole PHP
in the computer memory. It is easy to understand from Fig. \ref{DList}. Consider the suppression of the
$i$-th node. It consists in redirection of the upper in Fig. \ref{DList} pointer of the node $i+1$ to the
node $i-1$ and of the lower pointer of the node $i-1$ to the node $i+1$. The variables of the plug $i-1$ are
changed to account for the plug coalescence as described in sec. \ref{events} and their numerical meshes used
for liquid temperature calculation are merged. The $i$-th node can then be deallocated (i.e. the memory
occupied by it is liberated). These changes do not concern other nodes so they are not modified at all. The
standard array implementation would require the change of indices and a complete rewriting of the whole PHP
into the computer memory, which would slow down the execution because the corresponding amount of information
is quite large. For the closed loop PHP, the list is looped, i.e. a pointer belonging to the last node points
to the first node and vice-versa.

The PHP states at different time moments are recorded as another list. As previously, the standard array
cannot be used because the memory amount required for each PHP state may be different and is unknown in
advance. This ``PHP list'' needs to be only simply connected, which means that each node contains the PHP and
a (single) pointer to the next node. This pointer allows the sequential access to the PHP states, e.g. for
plotting. The PHP list may also be saved to a data file of a specific \texttt{.php} format that can than be
read by the postprocessing utility described below.

\subsection{General algorithm}

\begin{figure}[htb]
\centering
\includegraphics[width=0.8\columnwidth]{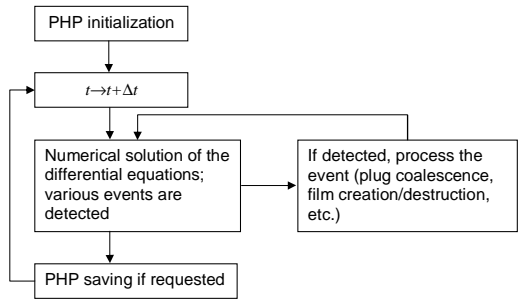}
\caption{General scheme of the C++ program.} \label{UML}
\end{figure}
The set of ordinary differential equations consists of eqs. (\ref{Lfl}, \ref{Lfr}, \ref{mv}, \ref{en},
\ref{Vl}, \ref{dml}, \ref{Vi}, \ref{hdd}), which totals to $\sum_i^M (2N_{e,i}+N_{l,i}+5)$ equations. Unlike
previous works \cite{shafii1,Sakul,Senjaya10}, they are written in the conventional differential form so that
their numerical integration can be performed with any numerical method (and not only explicit Euler method
used in \cite{shafii1,Sakul,Senjaya10}). The 4th order Runge-Kutta method (Fig. \ref{UML}) is used here. It
is renowned for its numerical stability and is thus better than the explicit method that can cause
oscillations of numerical origin. The evaluation of the right-hand sides of the equations may lead to a
detection of an event that changes the number of equations (see sec. \ref{events}). If detected, the event is
processed as discussed in sec. \ref{Data}. The number of equations and the equations themselves are updated
and the time step is recalculated.

\subsection{Data postprocessing}

\begin{figure}[htb]
\centering
\includegraphics[width=\columnwidth,clip]{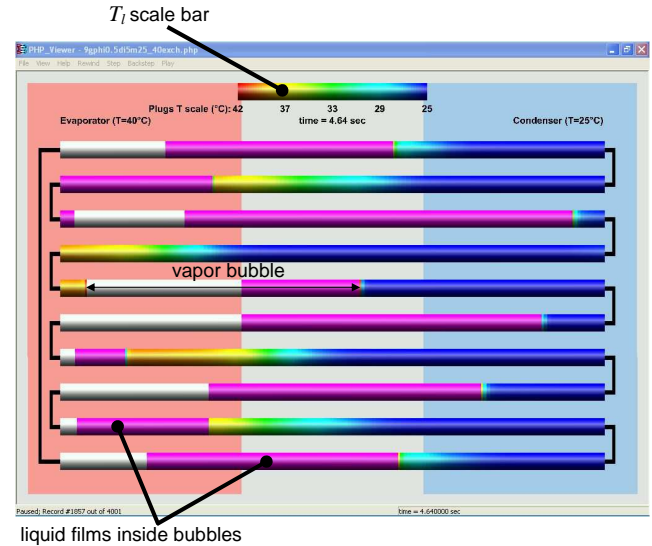}
\caption{(Color online) The screen of the PHP\_Viewer version 1.6 in the liquid temperature visualization
mode. The liquid films partially cover the internal tube walls inside the vapor bubbles. The temperature is
indicated with coloring of the liquid plugs.} \label{PHPViewer}
\end{figure}

The number of PHP variables is large and changes in time. The data analysis is impossible without clear
understanding of the position of each meniscus and film edge with respect to the PHP sections at each time
moment. In the absence of a suitable commercial software, a specific data postprocessing ``PHP\_Viewer''
utility had to be developed. PHP\_Viewer possesses a conventional Microsoft Windows graphic user interface
(Fig. \ref{PHPViewer}). The name of the \texttt{.php} file is displayed close to the application name at the
top of the screen above the menu. Some auxiliary information (the sequential number of the PHP record and the
corresponding time) is shown in the status bar at the bottom of the screen. The evaporator (left) and
condenser (right) locations are represented with rectangles, the width of which is to scale with respect to
the PHP (see Fig. \ref{Simu}). The condenser and evaporator temperatures are displayed above them. The time
of the current record is shown in between. The topology of the PHP bends is shown schematically with black
connectors. The tube diameter is not to scale. The liquid temperature in the plugs is represented with a
color. There are menu items usual for any video player. They allow controlling the animation speed and the
navigation inside the data file (stepping record by record, jumping to a record with a given number, etc.).

\section{Results and discussion}

The simulation runs were performed for the numerical parameters shown in Tab. \ref{param}. The initial
temperature of the fluid was chosen to be homogeneous and equal to $0.5(T_e+T_c)$. The equidistant bubbles are
distributed along the PHP. The menisci are initially at rest.
\begin{table}[htb]
  \centering
  \begin{tabular}{|c|c|}
    \hline
    \multicolumn{2}{|c|}{PHP parameters}\\
    \hline
    Fluid & water\\
    $N_p$ & 5 \\
    $d$ & 5 mm \\
   $L_e$ & 10 cm \\
   $L_c$ & 10 cm \\
   $L_a$ & 5 cm \\
    \hline
   \end{tabular} \begin{tabular}{|c|c|}
    \hline
    \multicolumn{2}{|c|}{Constants}\\
    \hline
    $\gamma$ & 0.47\\
    $L_m$ & 0.2 mm \\
    $U_m/U$ & $0.3$ \\
    $\Delta t$ & $10^{-4}$ s \\
    \hline
   \end{tabular}
  \caption{Parameters used for the numerical simulation.}\label{param}
\end{table}

\begin{figure}[htb]
\centering
\includegraphics[width=0.8\columnwidth,clip]{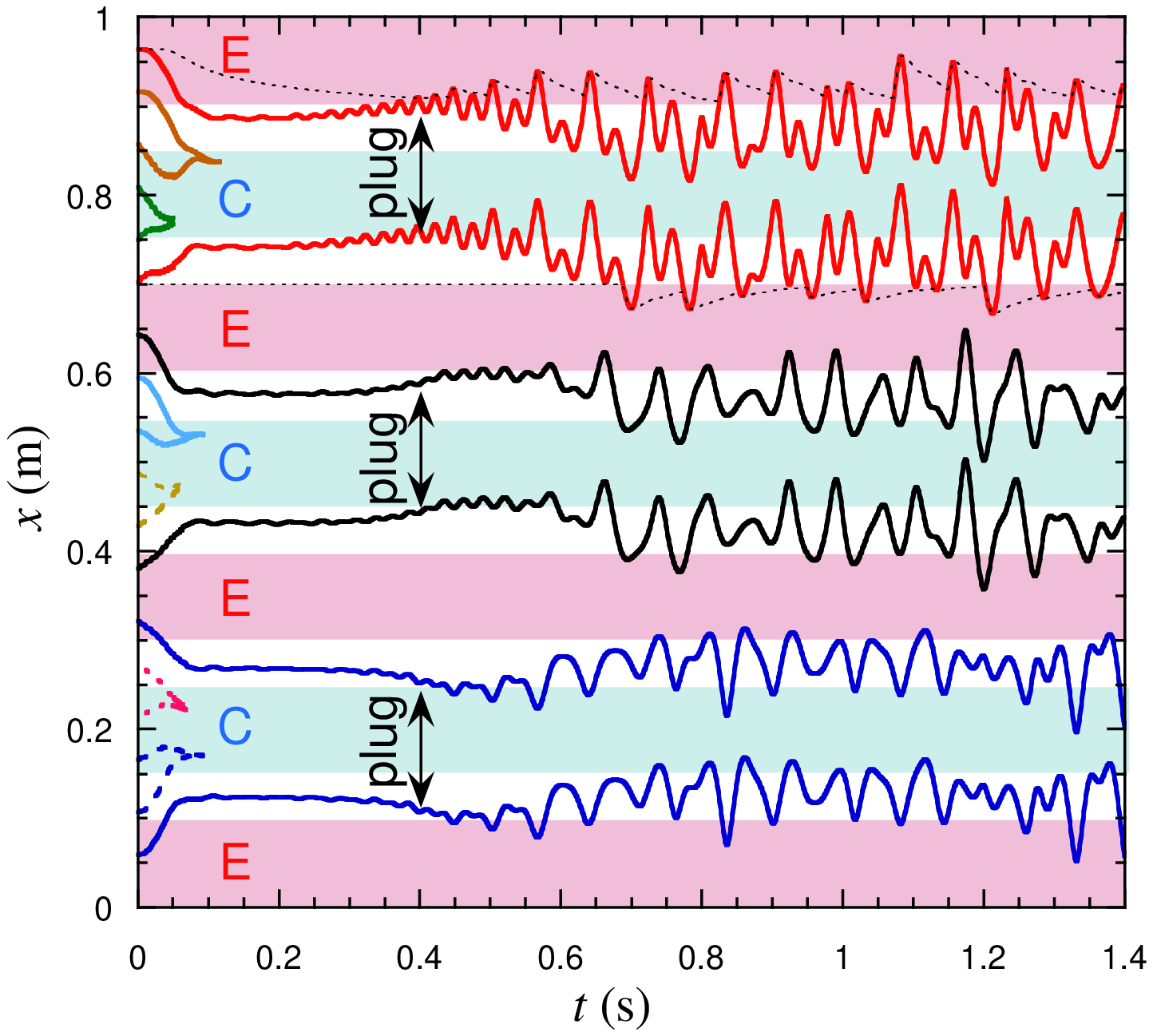}\\(a)\\
\includegraphics[width=0.8\columnwidth,clip]{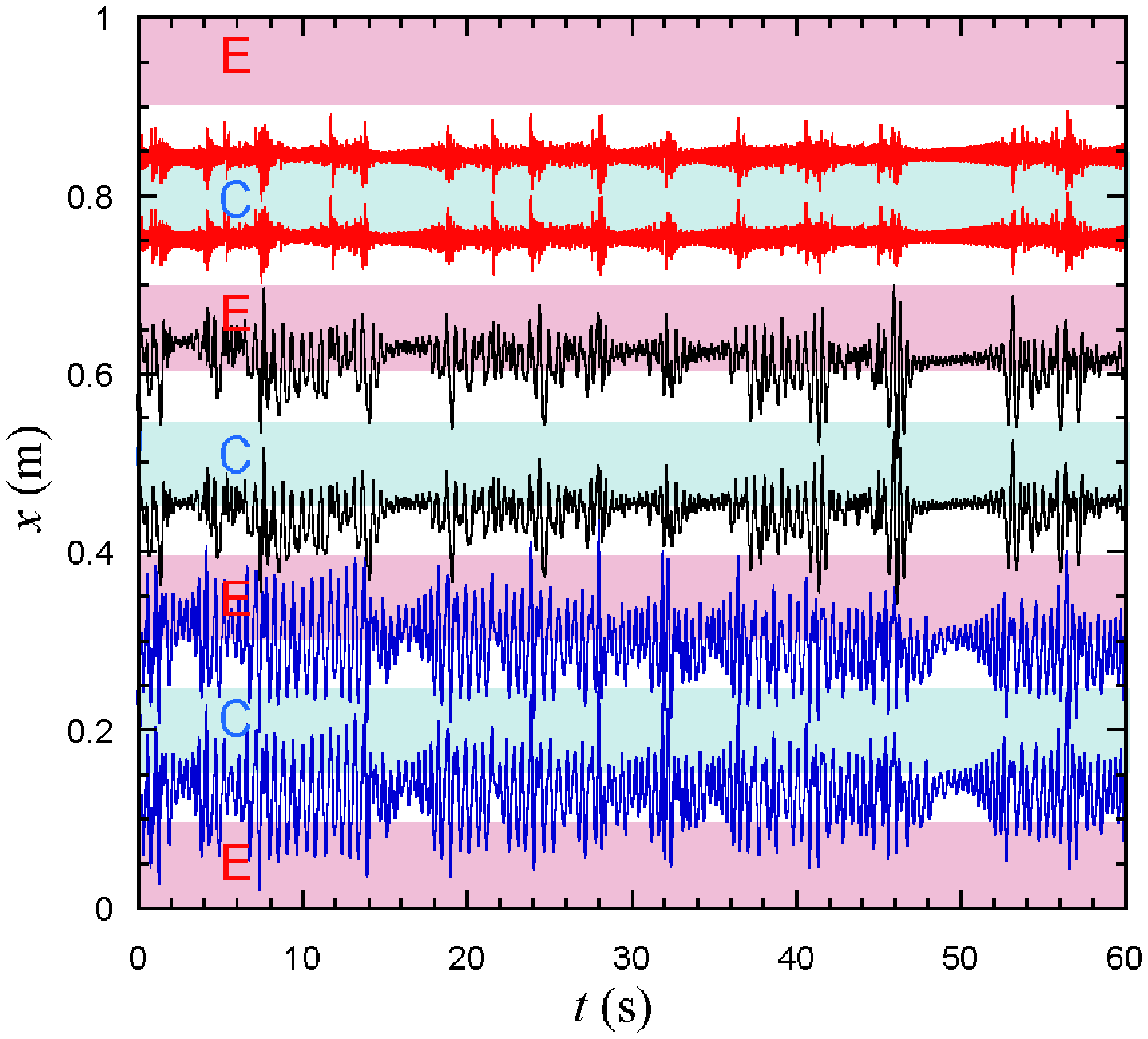}\\(b)
\caption{(Color online) Examples of the chaotic oscillation regime within the linear PHP representation
schematized in Fig. \ref{Simu}c. Evaporator (E) and condenser (C) sections are shown with the background
bars; adiabatic sections are in between. PHP is shown partially with evolution of only several menisci. (a)
Short time evolution. The bubble recondensation occurs at an early stage. The evolution of only two film
edges is shown by dotted lines. (b) Long time evolution for $T_e=45^\circ$ C, $T_c=25^\circ$ C, $\phi=0.55$,
$\delta_f=40\;\mu$m.} \label{chaot}
\end{figure}

It is well known \cite{Khan03} that there are many different regimes of PHP functioning. The present modelling
shows some of them, in particular the regime of chaotic oscillations (Fig. \ref{chaot}).
Its early stage is illustrated in Fig. \ref{chaot}a. Since the number of bubbles can only be reduced during
the PHP evolution, their large number (usually 9) is chosen initially. Fig. \ref{chaot}a shows that the
positions of some menisci join each other at $t<0.2$ s. This corresponds to the bubble recondensation that
occurs inside the condenser. The bubbles keep disappearing until only one per evaporator remains. The liquid
gathers in the condenser under the action of gravity (cf. Fig. \ref{Simu}a). The film that remains in the
evaporator does not, however, evaporate instantly (see the upper dotted line in Fig. \ref{chaot}a).  This may
cause an instability of the system, i.e. the development of oscillations. Their amplitude grows during a
short transient before attaining the developed oscillation regime. The amplitude in this regime depends on
the parameters (see below) and may be large. During large oscillations, the menisci penetrate both into the
condenser and the evaporator. The films persist in the evaporator; the film length oscillates (see the dotted
lines). The liquid volume change is almost invisible so that both ends of each liquid plug oscillate
synchronously. A portion of the $x$ axis corresponding to three plugs is shown in Fig. \ref{chaot}. They seem
to oscillate quite independently. Even the amplitude of their oscillations may be different: compare the
lower and upper plugs in Fig. \ref{chaot}b. The long-time PHP evolution (Fig. \ref{chaot}b) shows that the
oscillations are indeed chaotic: no periodic repetition can be mentioned. This is a dynamic chaos well known
to occur in the complex systems.

\begin{figure}[htb] \centering
\includegraphics[width=0.8\columnwidth,clip]{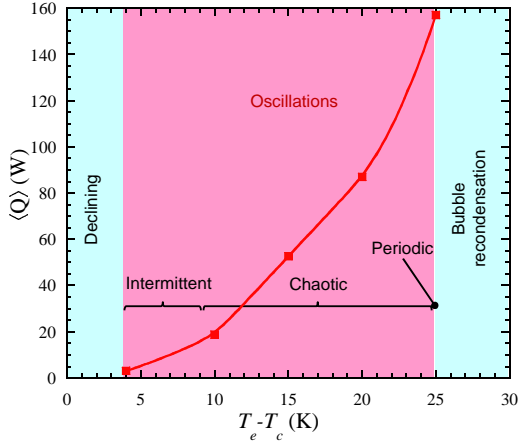}
\caption{(Color online) Heat transfer curve and PHP regimes for $T_c=25^\circ$ C, $\phi=0.55$,
$\delta_f=40\;\mu$m.} \label{Q-dT}
\end{figure}

The regimes of oscillations are convenient to be presented at the heat transfer curve (Fig. \ref{Q-dT}).
The oscillation exist when the temperature difference $\Delta T=T_e-T_c$ falls within an certain interval.
Within this interval, one may distinguish the chaotic regime discussed above and the intermittent regime. The
latter is characterized by a sequence of intervals during which the system oscillates strongly and the periods
of weak motion. Generally the amplitude is very small near the lower oscillation threshold. Below the threshold,
the oscillations decline to an equilibrium state where the vapor exists only in the evaporator and adiabatic
sections and the condenser sections are completely filled by the liquid. The pressure inside the bubbles becomes
equal to the saturation pressure corresponding to $T_e$, i.e. $T_e=T_{sat}(P_i)$ for every $i$. This state is
attained via condensation/evaporation, during which the mass of the vapor in each bubble relaxes to that
required by the vapor equation of state. The films in evaporator may exist but their lengths do not vary any
more (cf. Eq. (\ref{mfei}), the r.h.s. of which vanishes). At lower volume fractions, the liquid plugs do not
fill completely the condenser sections so that the above equilibrium state cannot be attained: the heat exchange
always exists. The stability of such a configuration is yet to be studied.

When $\Delta T$ is larger than the upper oscillation threshold, the initial perturbation eventually declines.
However the scenario is different from the low $\Delta T$ case. During the initial transient, the oscillations
develop. Their amplitude becomes large like in the chaotic regime; the bubbles are compressed strongly between
the plugs which have different inertia and thus move with different velocities. At some point one of the bubbles
is compressed so strongly that it recondenses. This leads to a creation of a liquid plug with yet larger
inertia, which causes the bubble recondensation in chain that in most cases ends up in a creation of a single
plug and a single bubble and the motion stops.

It is likely that the introduction of boiling will cause an instability of the final equilibrium state of the
bubble recondensation regime. Indeed, in the final single bubble state the liquid situates necessarily in the
evaporator and the boiling should occur and cause a restart of the oscillations.

The oscillation regime depends strongly on the chosen value of the film thickness $\delta_f$. The influence of
$\delta_f$ was studied for the following set of the fixed parameters: $T_e=35^\circ$ C, $T_c=25^\circ$ C,
$\phi=0.55$. It turns out that the $\delta_f$ decrease leads to the same sequence of regimes as $\Delta T$
growth. At $\delta_f>90\;\mu$m the self sustained oscillations are nonexistent. The $\delta_f$ decrease leads to
an appearance of the intermittent oscillations. Their amplitude grows as $\delta_f$ decreases until the bubble
recondensation appears and causes the oscillation disruption at $\delta_f\approx 5\;\mu$m. This shows the
importance of the $\delta_f$ choice.

\begin{figure}[htb]
\centering
\includegraphics[width=0.8\columnwidth,clip]{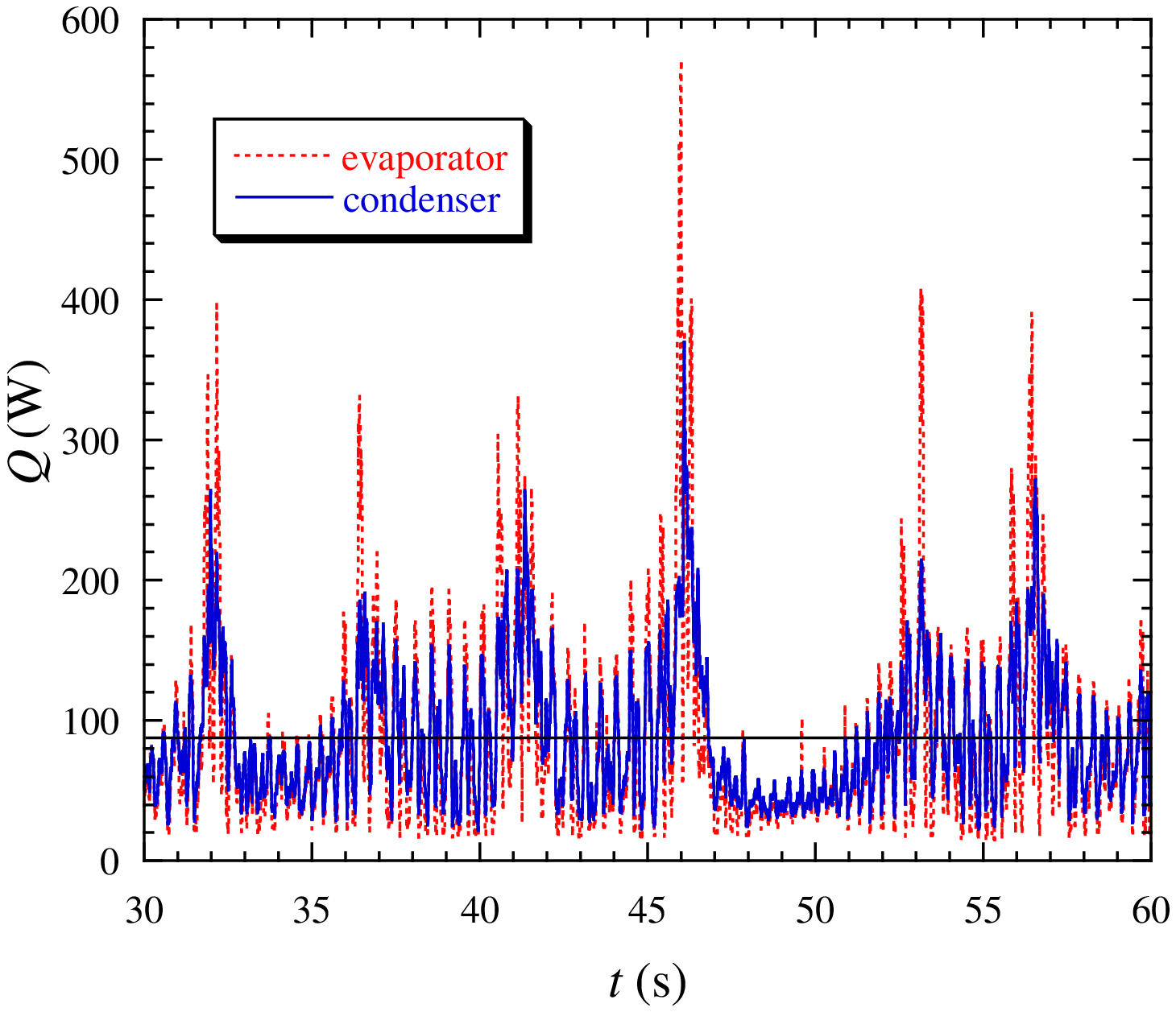}\\(a)\\
\includegraphics[width=0.8\columnwidth,clip]{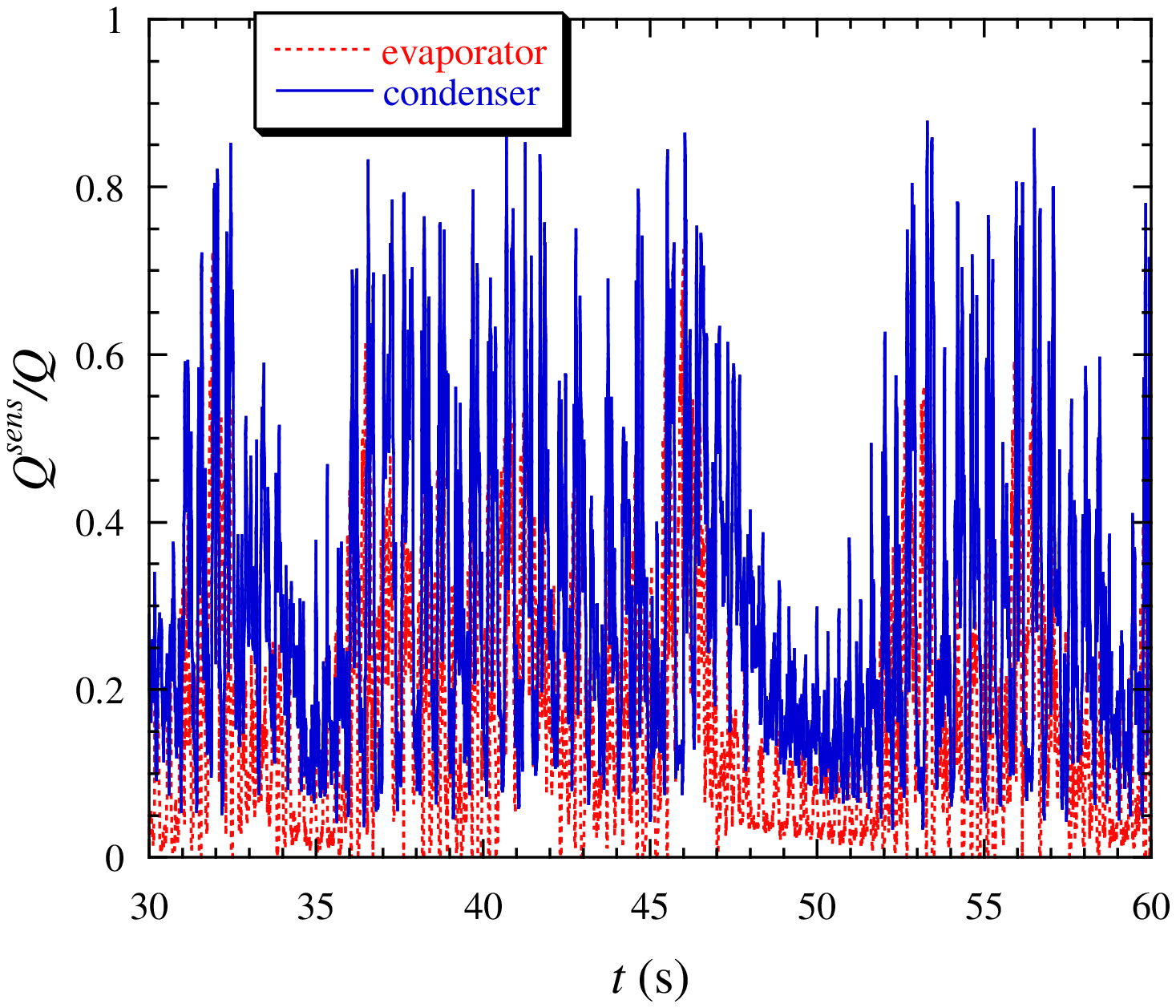}\\(b)
\caption{(Color online) Examples of the PHP heat exchange rate evolution in condenser and evaporator for the
same parameters as Fig. \ref{chaot}b. (a) Total heat exchange. The time average $\langle Q\rangle$ is shown
with a horizontal line. (b) Part of the sensible heat exchange. The average values are 0.18 for evaporator
and 0.29 for condenser.} \label{Exch}
\end{figure}

The heat transfer rate varies chaotically (Fig. \ref{Exch}a) during the oscillations accordingly to the
dynamics of the menisci.
During the developed oscillations, the equality (\ref{Qcond}) is satisfied within few per cent. The sensible
heat exchange part may however be different in the condenser and in the evaporator (Fig. \ref{Exch}b). It is
comprehensible since the condenser is occupied by the liquid most of the time. Accordingly, a part of the
sensible heat exchange in the condenser is larger than in the evaporator. The part of the sensible heat
exchange increases with the amplitude of the oscillations because the liquid sweeps more often hot and cold
walls. The averaged in time $Q$ value is shown in Fig. \ref{Q-dT} as a function of $\Delta T$. One can see
that quite efficient heat exchange can be achieved even without boiling that is likely to lead to the
continuation of the curve into the ``bubble recondensation'' region as discussed above.

\begin{figure}[htb]
\centering
\includegraphics[width=0.8\columnwidth,clip]{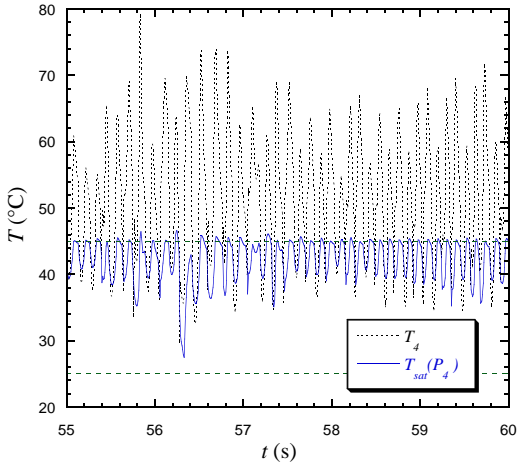}
\caption{(Color online) Typical evolution of temperatures obtained for a bubble from Fig. \ref{chaot}b. The
evolution of the left and right ends of this bubble correspond to the third and the second curves from the
top of Fig. \ref{chaot}b, respectively. The horizontal dashed lines correspond to $T_c$ and $T_e$.}
\label{Temp}
\end{figure}

The temperature inside the liquid is inhomogeneous, see Fig. \ref{PHPViewer}b. One can see the thermal
boundary layers that form near the menisci inside the liquid plugs. They appear because the pressure (and
thus the gas-liquid interface temperature $T_{sat}$) changes quickly during the oscillations; $T_{sat}$ is
sometimes 40-50 K larger than $T_e$. The analysis shows that the vapor pressures can also attain high values.
The vapor temperature rises strongly due to this compression and can be essentially higher than $T_e$ (Fig.
\ref{Temp}). This has been already observed in the single-bubble modelling \cite{IJHMT10}. During the
developed oscillations, the vapor is overheated: its temperature exceeds $T_{sat}$ by 10-20 K on average.
This shows that the hypothesis \cite{zhang-faghri,Holley05} about the vapor at saturation temperature is
hardly consistent.

The thickness of the boundary layers is different in different liquid plugs. It is defined by the value
$\sqrt{D\tau}$ where $\tau$ is an average period of the oscillations of a plug that grows with its mass.

\section{Conclusions}
\label{sec:conclusion}

A new model for PHP with arbitrary number of branches and arbitrary time-varying number of bubbles has been
presented above. It is more complex than the previous models and is capable of describing the chaotic
self-sustained oscillations of large amplitude. It is shown to reproduce correctly some features of experimental models like intermittent oscillation regime. Some analysis of the flow in the PHP and the heat transfer has been performed. An oscillation threshold occurs at small temperature difference. Another threshold, that occurs at a large temperature difference would probably be yet larger if the boiling were taken into consideration. The boiling thus needs to be implemented.

More studies need to be performed even for the present formulation of the model. In particular, the influence of the initial conditions (initial values of $M,T_i,X^s_i,$ etc.) might be of importance because the system is chaotic.

To perform more realistic modelling, more information is required on the phenomena that occur during the PHP
functioning. In particular, it is a priori evident that a strong viscous dissipation occurs in the liquid
films and near the contact lines (i.e. film edges). This effect leads to an additional pressure drop across
each meniscus. The available in the literature information on this phenomenon is scarce. The effect of the
PHP bends on the pressure drop should be accounted for. The film thickness is an important parameter, which
was imposed here to be constant and is taken to be micrometric like in previous works \cite{dobson1,
dobson2}. However it depends on the plug velocity and possibly on the evaporation/condensation rate and
should thus vary with time. Third, the vapor compression has not been yet assessed experimentally. It is not
clear if the liquid plug return force is caused entirely by the evaporation/condensation effect (assumed in
the models where the vapor was always at saturation temperature) or also by the vapor compression like in the
present approach.

\begin{acknowledgment}

Two of my students, B. Pottier and A. Tanniou, participated in the development of the simulation code during
their four month internship in our laboratory. Their contribution is gratefully acknowledged as well as the
corresponding CEA/SBT grants.
\end{acknowledgment}

\makeatletter\printnomenclature\makeatother
\appendix
\bibliographystyle{asmems4}
\bibliography{PHP,ContactTransf,Books}

\end{document}